\documentclass[a4paper,11pt]{revtex4}
\usepackage{graphicx}
\usepackage{epsfig}
\usepackage{dcolumn}
\usepackage{bm}
\usepackage{amsmath,amssymb,color}
\usepackage{setspace}
\begin{document}

\newcommand{\nwc}{\newcommand}
\nwc{\vs}{\vspace}
\nwc{\hs}{\hspace}
\nwc{\la}{\langle}
\nwc{\ra}{\rangle}
\nwc{\lw}{\linewidth}
\nwc{\nn}{\nonumber}
\nwc{\tb}{\textbf}
\nwc{\td}{\tilde}
\nwc{\Tr}{\tb{Tr}}
\nwc{\dg}{\dagger}

\nwc{\pd}[2]{\frac{\partial #1}{\partial #2}}
\nwc{\zprl}[3]{Phys. Rev. Lett. ~{\bf #1},~#2~(#3)}
\nwc{\zpre}[3]{Phys. Rev. E ~{\bf #1},~#2~(#3)}
\nwc{\zpra}[3]{Phys. Rev. A ~{\bf #1},~#2~(#3)}
\nwc{\zjsm}[3]{J. Stat. Mech. ~{\bf #1},~#2~(#3)}
\nwc{\zepjb}[3]{Eur. Phys. J. B ~{\bf #1},~#2~(#3)}
\nwc{\zrmp}[3]{Rev. Mod. Phys. ~{\bf #1},~#2~(#3)}
\nwc{\zepl}[3]{Europhys. Lett. ~{\bf #1},~#2~(#3)}
\nwc{\zjsp}[3]{J. Stat. Phys. ~{\bf #1},~#2~(#3)}
\nwc{\zptps}[3]{Prog. Theor. Phys. Suppl. ~{\bf #1},~#2~(#3)}
\nwc{\zpt}[3]{Physics Today ~{\bf #1},~#2~(#3)}
\nwc{\zap}[3]{Adv. Phys. ~{\bf #1},~#2~(#3)}
\nwc{\zjpcm}[3]{J. Phys. Condens. Matter ~{\bf #1},~#2~(#3)}
\nwc{\zjpa}[3]{J. Phys. A: Math theor  ~{\bf #1},~#2~(#3)}

\newcommand\bea{\begin{eqnarray}}
\newcommand\eea{\end{eqnarray}}
\newcommand\beq{\begin{equation}}  
\newcommand\eeq{\end{equation}}
\newcommand{\new}{\newpage}
\newcommand{\noi}{\noindent}
\newcommand{\bib}{\bibitem}
\newcommand{\cosec}{\operatorname{cosec}}
\newcommand{\non}{\nonumber}  
\newcommand\mb{\mathbf}
\newcommand\bs{\boldsymbol}
\newcommand\mT{\mathcal{T}}
\newcommand\dd{\text{d}}
\newcommand\s{\sigma}
\newcommand{\p}{\tilde{\Psi}}
\newcommand{\ps}{\tilde{\Phi}}
\newcommand{\C}{\tilde{c}_{j}}
\newcommand{\X}{\tilde{\chi}_{{\bs k}}}
\newcommand\ie{{\it{i.e.}}}
\newcommand\etal{{\it{et al.}}}
\newcommand\eg{{\it{e.g.}}}\def\bbraket#1{\mathinner{\langle\hspace{-0.75mm}\langle{#1}\rangle\hspace{-0.75mm}\rangle}}
\def\i{\imath}
\def\v{\upsilon_F} 
\def\nn{\nonumber}
\def\f{\frac}
\def\al{\alpha}
\def\om{\omega}
\def\de{\delta}
\def\ep{\epsilon}
\def\ga{\gamma}
\def\si{\sigma}
\def\Do{\partial}
\def\De{\Delta}
\def\mb{\mathbb}
\def\mc{\mathcal}
\def\vr{\varrho}
\def\d{\cdot}
\def\t{\tilde}
\def\l{\lambda}
\def\la{\langle}
\def\ra{\rangle}
\def\mbb{\mathbb}
\def\Y{\Upsilon}
\def\ua{\uparrow}
\def\da{\downarrow}
\def\sf{\textsf}
\def\al{\alpha}
\def\be{\beta}
\def\til{\tilde}
\def\ka{\kappa}
\def\sX{\small{X}}
\def\br{{\bs r}}
\def\bk{{\bs k}}

\title{How long does a quantum particle or wave stay in given region of space?}

\author{S. Anantha Ramakrishna$^1$  and Arun M. Jayannavar$^{2, 3}$}

\address{$^1$ Department of Physics, Indian Institute of Technology Kanpur, Kanpur 208016 \\ $^2$Institute of Physics, Sachivalaya Marg, Bhubaneswar 751005, India.
\\$^3$Homi Bhabha National Institute, Training School Complex, Anushakti Nagar, Mumbai 400085, India.}

\begin{abstract}
The delay time associated with a scattering process is one of the most important dynamical aspects in quantum mechanics.  A common measure of this is the Wigner delay time based on the group velocity description of a wave-packet, which my easily indicate superluminal or even negative times of interaction that are unacceptable.  Many other measures such as dwell
times have been proposed, but also suffer from serious deficiencies, particularly for evanescent waves.  One important way of realising a timescale that is causally connected to the spatial region of interest has been to utilize the dynamical evolution of extra degrees of freedom called quantum clocks, such as the spin of an electron in an applied magnetic field or coherent decay or growth of light in an absorptive or amplifying medium placed within the region of interest. Here we provide a review of the several approaches developed to answer the basic question “how much time does a quantum particle (or wave) spend in a specified region of space?”  While a unique answer still evades us, important progress has been made in understanding the timescales and obtaining positive definite times of interaction by noting that all such clocks are affected by spurious  scattering concomitant with the very clock potentials, however, weak they be and by eliminating the spurious scattering.
\end{abstract}

\maketitle




\section{Introduction}
One of the first things a student of Physics learns is to calculate the time at which an event, such as the location of a particle at a given position, occurs. A related question would be to ask how long does a particle stay in a given region of space during the course of its motion. Classically it is possible to simultaneously specify the position and the momentum of a particle, and these questions about the time instants or the time intervals for given events have unambiguous answers.  Even as we proceed to statistically understand systems with very large number of particles through probability distributions, where the classical motion of each particle is described only stochastically, concepts such as the first passage times~\cite{ref_classical_first_passage} remain meaningful.  This well entrenched concept, however, becomes difficult to define for quantum mechanical systems, and indeed,  for any form of  waves. 
\begin{figure}[!thpb]
\includegraphics[width=0.48\linewidth]{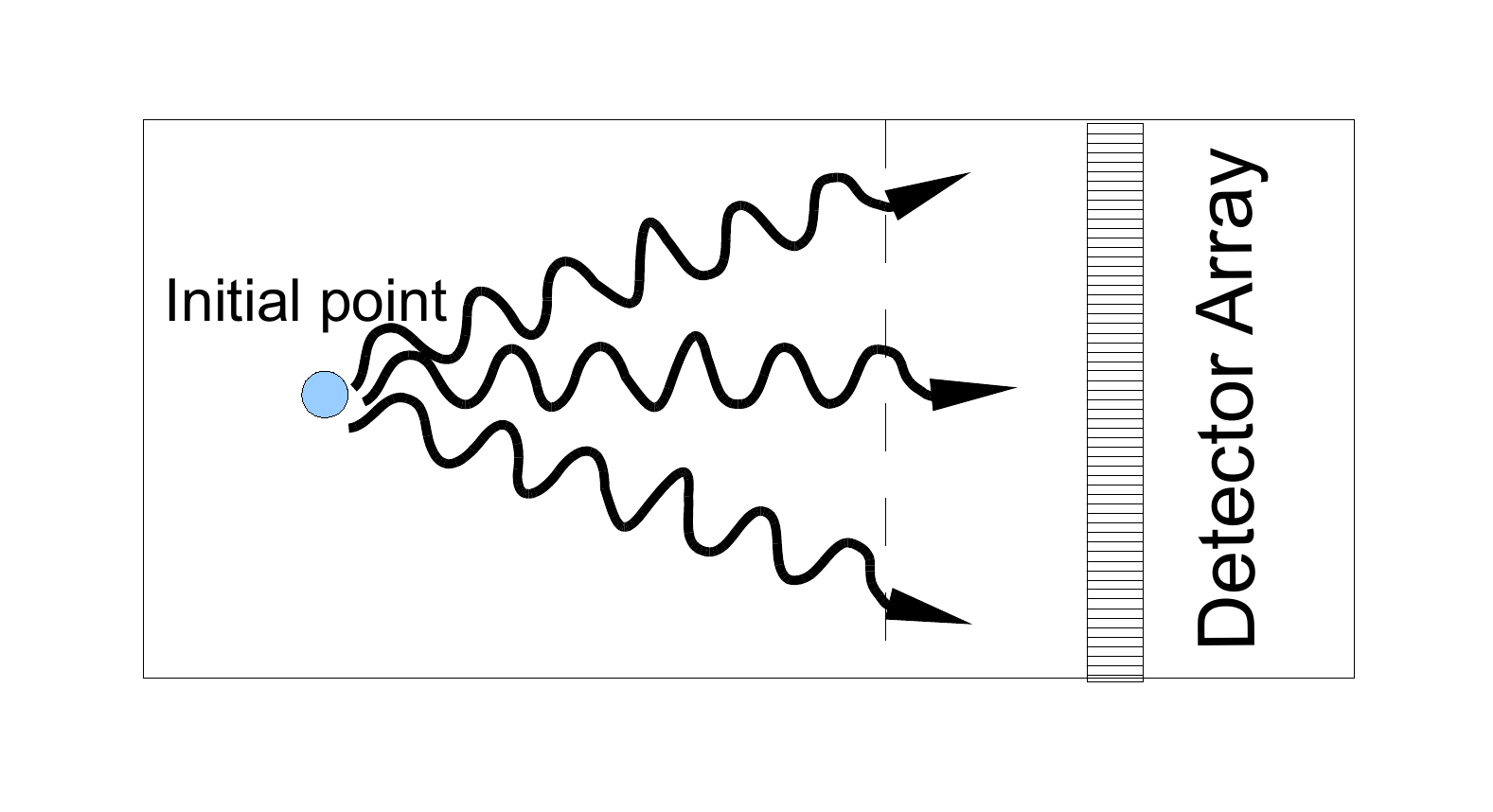} \includegraphics[width=0.48\linewidth]{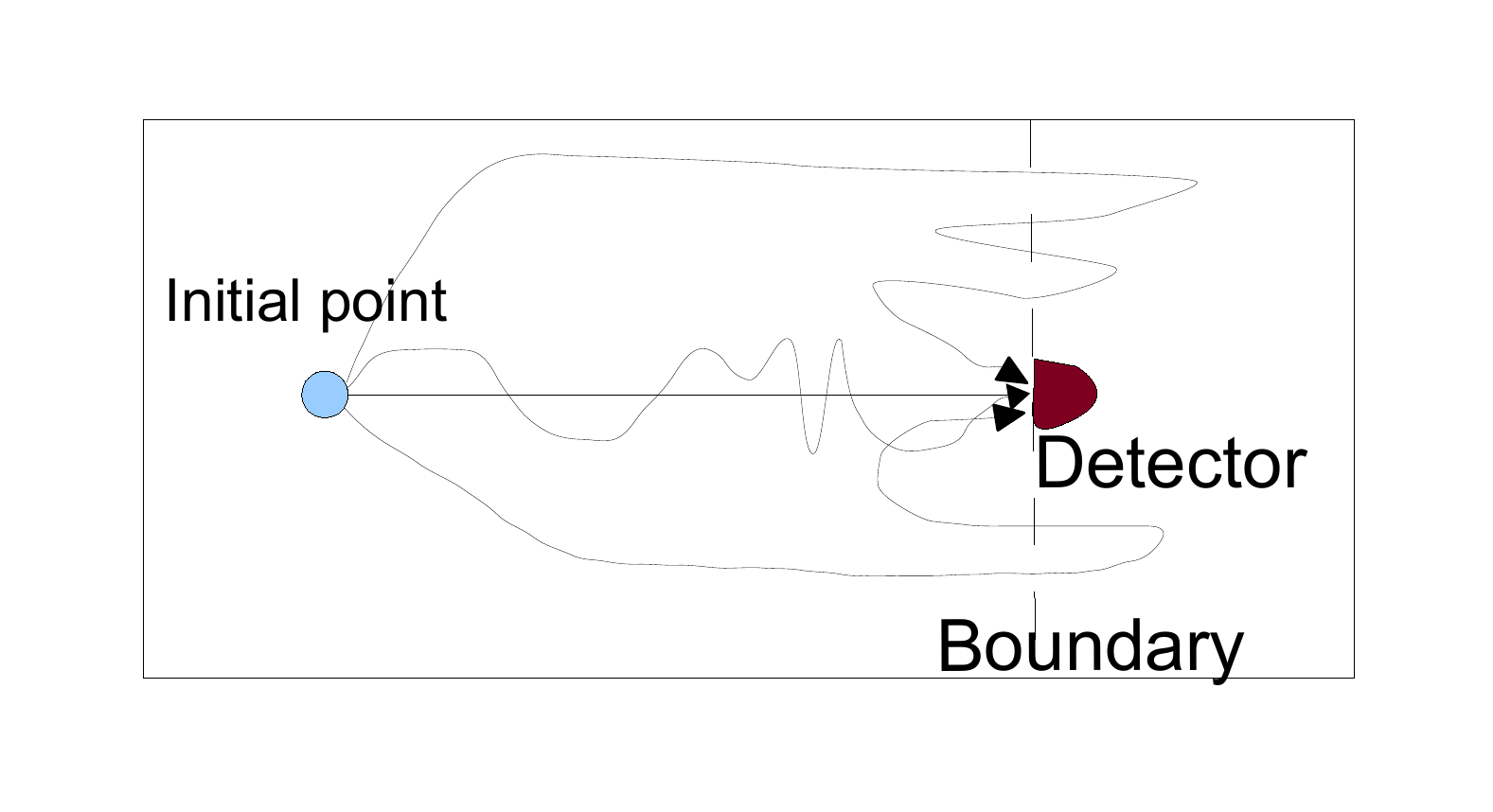}
\caption{Left panel: Schematic picture of a quantum particle that is initially localized in a fixed region and arrives at a detector array. The first passage time refers to the first point in time when the particle is registered at a detector. Right panel: The particle may take any virtual path to 
get to the detector, including ones that cross the boundary many times. The final amplitude will be a sum over the amplitudes for any such path. }
\label{fig1}
\end{figure}

	Imagine an experiment where a quantum-mechanical particle is released from some fixed region inside a box. On one side of the box there is a screen with detectors which click as soon as the particle "arrives" at the screen.  One expects that the time of arrival of the particle is a stochastic variable and it is interesting to ask for it's probability distribution. This is similar to asking for the distribution of the time of absorption of a Brownian particle at some point. However, the quantum problem turns out to be very subtle and there is as yet no clear answer to the question. The point is that the particle that is initially localized and released must subsequently cross a given boundary in space when it is detected for the very first time. Yet this very calculation includes amplitudes from paths that may very well extend all over space even beyond the given boundary. Very recently the problem of  first time arrival of a quantum mechanical particle has been considered satisfactorily utilizing   a  path integral approach that with a restricted path decomposition and appears to succeed in obtaining positive definite quantum first passage times for motion on a one dimensional lattice~\cite{nkumar2014}. 

The principal difficulty arises from the fact that waves are infinitely deformable objects and many aspects of motion arise through interference effects.  Hence it becomes impossible to define well defined start and finish lines for wave packets. In the quantum mechanical perspective, where all dynamical observables have a corresponding operator, there has been great difficulty in defining a “good” Hermitian operator that conforms to classical notions of time or a time interval~\cite{ref_times_operator}. For classical waves such as electromagnetic waves also, related difficulties exist. For example, we would like to classically demand that a time of stay in a given region of space should be (i) Real, (ii) Positive definite, (iii) tend to classically calculable times in the limit of large energies. Inspite of serious efforts over many years~\cite{landauer_review}, no definition for such a quantum mechanical time has been universally accepted. In fact, when we say that a wave moves, it becomes imperative to clearly define what is the quantity related to the motion of a wave that is being talked about. Regardless of these difficulties, it must be emphasized that the times of traverse or dwell associated with a wave are calculable quantities that can be useful for understanding processes and process rates in a given system.  

In this tutorial, we will discuss the several approaches, the related difficulties and some recent possibilities that have arisen in this context.  First, we will go through some of the basic definitions and concepts regarding wave motion and the times that can be related to the motion of a wave.  It turns out that there is a large variety of times arising from different definitions related to various different quantities associated with a wave.  Surprisingly, these definitions can result in apparent superluminal times or even negative delay times for the transit of a wave through dispersive media or potentials. Subsequently,  we will describe certain clocking mechanisms associated with physical processes such as the precession of a spin in a magnetic field~\cite{buttiker_larmor} that have been developed to calculate theoretically the traversal or dwell times of a wave~\cite{landauer_review}.  These clocking mechanisms have to be used with care as they can contribute to scattering and change the very problem being discussed~\cite{sar_epl}.  These ideas lead to the possibility of using quantum dephasing or stochastic absorption as a clock. 

\section{1. Time-scales based on the group velocity {\label{sec:II}}}
Here we will try to explore these questions in the general context of waves to cover both electromagnetism and quantum mechanics. Both the time independent Schrodinger equation  for the wavefunction of  a quantum particle and the Maxwell equations in frequency domain for the amplitude of a time harmonic electromagnetic wave (of only one polarization) reduce to the Helmholtz wave equation:
\begin{equation}
\nabla^2 \psi + k^2 \psi = 0.
\end{equation}
For the electromagnetic wave, k is the wave vector given by $k^2 = n^2 \omega^2 /c^2$,  $\omega$ is the angular frequency of the wave and $n$ is the refractive index of the medium with $c$ being the speed of light in vacuum. For the quantum particle in comparison, $k$
is given by $k^2 = 2m(E-V_0)/\hbar$ , where$E$ and $m$ are the energy and the mass of the particle and$V_0$ is the potential. The principle difference between the the two systems is in free space where the refractive index is unity. Consequently the wave-vector for an electromagnetic wave is linearly proportional to the frequency. In the case of the quantum particle, even for $V_0=0$, the wave-vector
disperses with the quadratic root of the energy. This has a non-trivial manifestation in the undistorted propagation of electromagnetic pulses in free space while a quantum mechanical wavepacket spreads out and disperses even in free space. Thus, it is realized that the fundamental properties of propagation of the waves are governed by the potentials or the refractive index of the medium through the {\it dispersion relation} between the wave-vector and the frequency or energy.

Consider the scalar wave $\psi(\vec{r},t) = a(\vec{r}) \exp[i\phi(\vec{r}) - i\omega t]$  where $\phi(\vec r)$  is some scalar function. For this wave,  $\phi(\vec r) = \mathrm{constant}$ denotes the constant phase surfaces. To trace the motion of these surfaces, let us look at the condition at two points $(\vec{r},t)$ ,  and $(\vec{r} + \delta \vec{r}, t+\delta t)$. The phase front is the same if, and only if,
\begin{equation}
\phi (\vec{r} + \delta \vec{r}) -\omega( t+\delta t)  \simeq \phi(\vec{r}) + \delta \vec{r} \cdot \nabla \phi(\vec{r}),t) - \omega t - \omega \delta t = \phi(\vec{r})  - \omega t, 
\end{equation}
where we have included the first order term only in the infinitesimal $ \delta \vec{r}$.  FRom the above, we obtain 
\[ v_p  = \left| \frac{\delta \vec{r}}{\delta t} \right| = \frac{\omega}{| \nabla \phi |} \]
as the phase velocity for a wave with an arbitrary wave front. The ratio , where $\phi(\vec r)$ is the phase of the wave~\cite{born_wolf}, is known as the phase velocity of the wave and represents the rate at which the equiphase surfaces of the wave move through the medium. Note that the phase velocity can be just about any number (positive or negative) depending on the phase structure (gradient) of the wave. In one dimension or when there is transverse invariance along two dimensions and plane waves result, the phase velocity reduces to the familiar relation. This gives rise to the conventional notion that the phase velocity for a wave is $c/n$, and is hence mistaken for the speed of a wave in a medium. As pointed out, the phase velocity can easily be superluminal for materials with refractive index $n < 1$. Further, in negative refractive index materials~\cite{sar_book} and in waveguides that support backward wave propagation, this phase velocity is obviously negative. Thus, the phase velocity does not really specify anything concrete about the rate of the motion of the wave or the time spent by a wave in a given region of space. 

In a material medium, the refractive index (polarization) of a medium will, in general, be frequency dependent and a complex quantity. This is a simple consequence of the medium having certain natural frequencies at which it will resonantly polarize corresponding to atomic or molecular levels of the constituents of the medium. Due to the different refractive index at different frequencies, the  constituent time harmonic waves present in a wave packet essentially travel at their own phase velocities resulting in the interference pattern changing completely in time and leading to a dispersion of the wave packet. This is easily seen by writing down the field amplitude of the wave at different times. If $E(\vec{r},0)$ be the field amplitude at time$t =0$, and 
\begin{equation}
E(\vec{k} = \int_{-\infty}^\infty E(\vec{r},0) e^{-i\vec{k}\cdot\vec{r}} ~d^3r
\end{equation}
is is the spatial Fourier transform of the field amplitude, the field amplitude at any other time, $t$, is given by 
\begin{equation}
E(\vec{r}, t) = \left(\frac{1}{2\pi}\right)^3  \int_{-\infty}^\infty E(\vec{k}) e^{i(\vec{k}\cdot\vec{r}-\omega t)} ~d^3k .
\end{equation}
If the dispersion was linear, this would just correspond to the same function
shifted to a new position. For an arbitrary dispersion $\omega(\vec{k})$, it becomes difficult to say anything in general. If we assume, however, that most of the amplitude is concentrated in a small frequency band about a central (carrier) frequency $\omega_0$, then one can carry out a Taylor’s series expansion
\[ \omega(\vec k) = \omega(\vec k_0) + (\vec {k} -\vec{k}_0) \cdot \nabla_k \omega(\vec{k}_0) + \cdots , \]
where the subscript $k$ indicates that the derivative is with respect to the wave-vector and retain only the linear term. Substituting this in the expression for the field amplitude, one obtains, 
\begin{equation}
E(\vec{r}, t) = \left(\frac{1}{2\pi}\right)^3  e^{i\varphi} \int_{-\infty}^\infty E(\vec{k}) e^{i\vec{k}\cdot[\vec{r} - \nabla_k\omega(\vec{k}_0) - \omega t]} ~d^3k ,
\end{equation}
which is essentially the same waveform that is shifted by an amount $\nabla_k \omega(\vec{k}_0)t$ in space, apart from a trivial extra phase of $\varphi = \omega_0 t + k_0 \cdot \nabla_k \omega(\vec{k}_0)t$. This brings up another rate at which the envelope of the wave propagates in the medium and defines the so-called {\it group velocity} $v_g = \nabla_k \omega(\vec{k}_0)$. This tracks the rate at which fiducial (well recognisable) points on the waveform, such as the peak of the wavepacket, move and it is assumed that waveform is largely undistorted. This is rarely satisfied for a quantum mechanical wave even in free space. For an electromagnetic wave, however, this is satisfied for quasi-monochromatic fields and when absorption in the medium is minimal. Again, it should be noted that the spread in the wavevectors is also equivalently small and the superposition should consist of waves moving along a common direction.

Noting that the group delay time accumulated upon traversing a distance $L$ in  the medium is
\[ \frac{L}{\nabla_k \omega(\vec{k}_0)} = \left. \frac{\partial (kL)}{\partial \omega}\right|_{\omega_0} ,\]
one can equivalently define a group delay time for scattering problems where the kinematic phase is replaced by
the phase change upon scattering ($\phi$). In one dimension, this would be the phase shift upon reflection or transmittance. This yields the {\it Wigner’s group delay time} 
\begin{equation}
\tau_w = \left. \frac{\partial \phi}{\partial \omega} \right|_{\omega_0}.
\end{equation}
Note the extrapolation made here from a free propagation problem into the time delay obtained in a scattering event. This time essentially measures the time difference between the entry and emergence of fiducial points into and out of
the scattering volume.

\begin{figure}[t]
\includegraphics[width=0.9\linewidth]{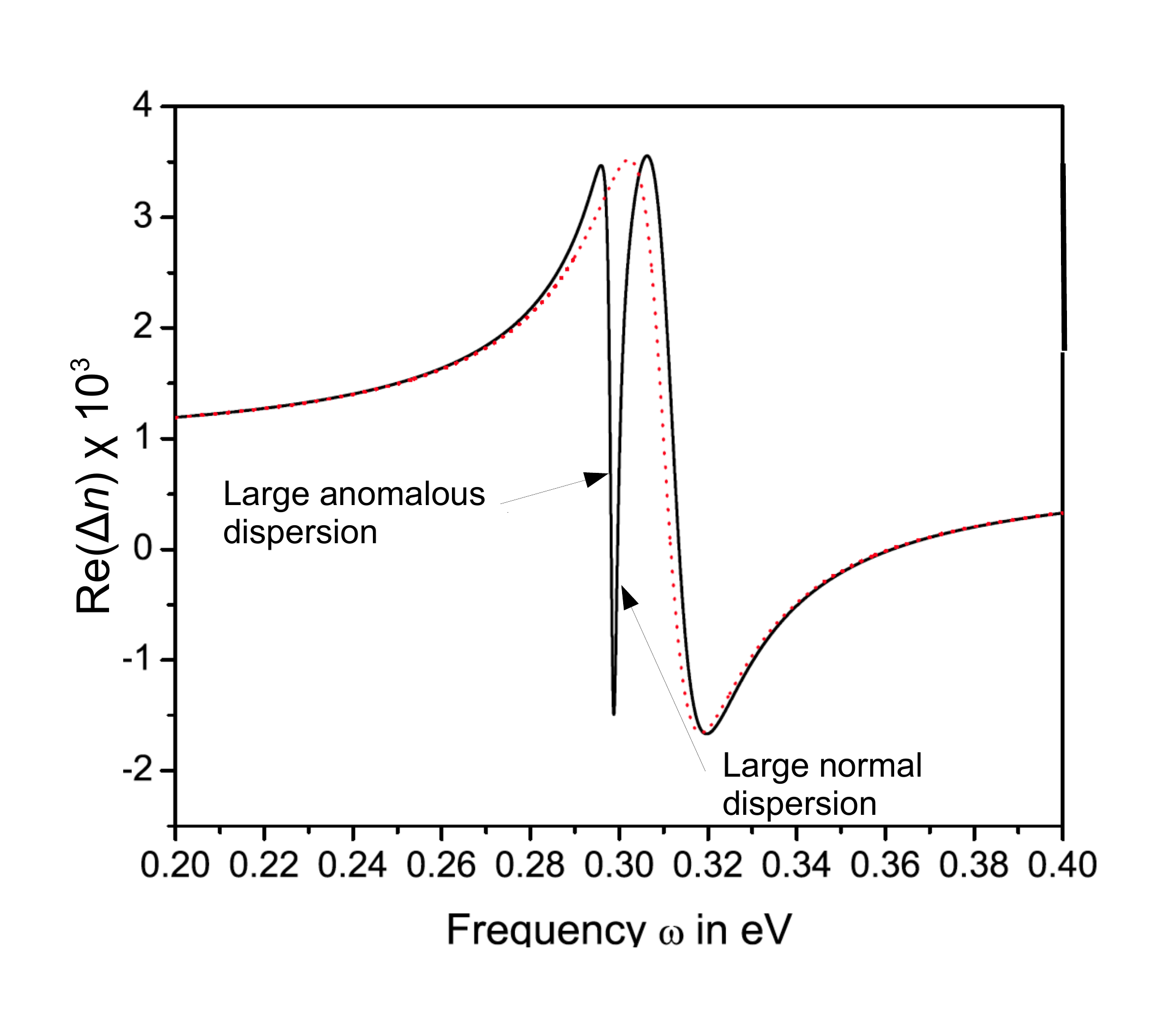} 
\caption{The change in refractive index due to dispersion is plotted with a magnified scale. The dispersion of the real part of the refractive index for a single Lorentz resonance (red dotted curve) and for a highly dispersive medium that can be produced, for example by electromagnetically induced transparency or by a Fano resonance.  Such dispersions enable very large changes of the refractive index, either normal or anomalous as marked in the figure. The group velocity can easily become very small or very large and even negative in such dispersive regimes.}
\label{fig2}
\end{figure}

Using the dispersion for light,  $k^2 = n(\omega)^2 \omega^2/c^2$in a homogenous isotropic medium with a dispersive refractive index $n(\omega)$ , the group velocity is also often conveniently written in the following manner
\begin{equation}
v_g = \frac{c}{n(\omega_0) + \omega_0 (dn/ d\omega)_{\omega_0}}.
\end{equation}
The group velocity from this definition immediately shows that the group velocity can be very small when the dispersion is normal and large, i.e., $(dn/d\omega) \gg 0$  (see Fig.~\ref{fig2}). This is the origin of ultra-slow light which has been demonstrated using the large dispersion possible in atomic gas vapours and Bose-Einstein condensates~\cite{ultra_slow_light}. On the other, if the dispersion were anomalous and large , $(dn/d\omega) < 0~\mathrm{and}~ |dn/d\omega| \gg 0$ (see Fig.~\ref{fig2}), then we have that the group velocity can become greater than c, or even negative in such situations. Indeed there have been several experiments on the apparent motion of light in a medium at a rate faster than light in vacuum [REF]. Negative group velocities and negative Wigner delay times are even more problematic – that would imply that the pulse exited region of space before it even entered, in a sense violating causality. 

Landauer~\cite{landauer_review} had severely criticised the Wigner delay time to give information about the dwell times on grounds of causality and emphasized that there was no causal connection between the peaks (or fiducial points) of the incoming and the outgoing waveforms. This difficulty is accentuated when the scattering potential strongly deforms the wave-packet. The large dispersion responsible for the deformation is also usually accompanied by severe dissipation or gain that causes large spectral modifications with corresponding distortions of the temporal pulse shape. Hence the analysis of the motion of a spectrally broad or distorted pulse in terms of the conventional group velocity is severely limited, as the pulse can lose its very identity after propagating to a large distance in the dispersive medium. These arguments apply and hold true even for definitions of the times based on the center-of-mass of the wavepacket or in an alternative approaches where the motion of the forward edge of the wavepacket is followed. 

These issues raise questions about the group velocity, or the Wigner delay time that is based on the group velocity, to actually provide an answer to the question that we seek about the time that the wave spends in a given region of space.  It has been emphasized by most authors  that the superluminal or negative Wigner delay times do not violate causality or the Special theory of relativity. It simply turns out that the functions that we considered to describe the fields have been smooth analytic functions. This is a consequence of the fields being the solutions of the Helmhotz differential equation. Analytic functions are problematic because they have an infinite support – the function extends all over the space, although it may be very small. A good example of such functions is the Gaussian function. In principle, one may always take the analytic function at a given point and carry out a Taylor's series expansion of the function using the values of the function and its derivatives at the given point, to obtain its value at any point in space, however, far off from that point. Since the function is analytic, all the derivatives at the given point exist. Thus, knowledge of the analytic signal even if localized in the manner of a Gaussian function, already exists at all the other points in principle, even if the value of the signal were to be infinitesimally small. Thus, there is no extra information being conveyed to the other points with the wave motion as the information was already present. Thus, there is no violation of the special theory of relativity or of causality.  It has been emphasized that in principle only meromorphic functions (with discontinuities in the function or its  derivatives) can be used to encode information. Any such discontinuity will generate very high frequency components in the power spectrum of the signal. These high frequency components will always propagate at the speed of light in vacuum (c) as the refractive index as This is a consequence of finite inertia for the charge carriers in all material media.

\section{The Dwell times based on current fluxes or energy transfer{\label{sec:III}}}
A second approach to this problem has been to define the time in terms of probability of finding the particle within the spatial volume of interest. Smith~\cite{smith_dwell_times} defined the  dwell time for a mono-energetic  quantum particle in the region $[0, L]$  (in one dimension) as 
\begin{equation}
\tau_d = \frac{1}{J} \int_0^L \lvert \psi(x) \rvert^2 ~dx
\end{equation}
where $\psi(x)$ is the wavefunction and $ J = \mathrm{Re} (\hbar/im) (\psi^*\;\nabla \psi)$ is the current flux associated with the incoming particle. Note that the {\it Smith dwell time},  is independent of the scattering channel (reflection or transmission) and is hence, an unconditional time.  In case of a time-varying pulsed waveform, this dwell time can be generalized by integrating over all times  as
\begin{equation}
\tau_d = \int_{-\infty}^{\infty} dt ~ \int_0^L \lvert \psi(x) \rvert^2 ~dx.
\end{equation}

In the case of electromagnetic waves, a similar approach can be adopted, but using the energy of the fields. Thus the dwell time in a region of volume $(V)$ could be defined as  
\[ \tau_d = \frac{\int_V U d^3r}{ \int_A \vec{S} \cdot d\vec{a}} , \]
where $U$ is the energy density associated with the electromagnetic wave, $\vec{S}$ is the Poynting vector denoting the power flow per unit area of the incoming wave and A is the surface area of volume through which the incoming wave is incident. While there is no problem with the Poynting vector defined as $\vec{S} = \vec{E}\times \vec{H}$ in terms of the electric and magnetic fields, there are severe difficulties in defining an energy density solely associated with the wave in a dispersive and dissipative (or amplifying) medium. The difficulties of separating the energy associated with the wave and the polarization in the medium are well known in this scenario where the two fields continuously exchange energy. In fact, it is well known that the energy density defined by first order Taylor expansions of the dispersion~\cite{landau_lifschitz} can easily be negative for severe dispersion of the material parameters.  In some sense, the above quantity  and the Smith Dwell time are equivalent as both the quantities involve quadratic expressions of the underlying fields. The issue of dispersion and polarization in material media complicate the issues further in the context of electromagnetic waves.  

Another fruitful approach to define arrival times that is based on the centroid of  power flow of the electromagnetic wave is noteworthy~\cite{peatross}. An arrival time at a point can be defined as a time average (first moment of time) over the component of the Poynting vector normal to a surface
\begin{equation}
\langle t \rangle_r = \frac{\hat u \cdot \int_{-\infty}^\infty t \vec{S}(\vec{r},t) dt}{ \hat u \cdot \int_{-\infty}^\infty \vec{S}(\vec{r},t) dt},
\end{equation}
where $\hat u$ denotes the unit normal to the given surface, which could very well be that of a detector. The time for traverse between two points $\vec{r}_i$ and  $\vec{r}_f$ can now be thought of as the difference between the arrival times at the two points as $\Delta t = \langle t\rangle_{\vec{r}_f} -  \langle t\rangle_{\vec{r}_i}$. A basic theorem was proven~\cite{Peatross} to show that the propagation delay could be decomposed in terms of a net group delay and a reshaping delay. The net group delay consists of a spectrally wieghted average group delay at the final point $\vec{r}_f$  given by 
\begin{equation}
\Delta t_G = = \frac{\hat u \cdot \int_{-\infty}^\infty t \vec{S}(\vec{r}_f,\omega) [(\partial \mathrm{Re}(k) / \partial \omega) \cdot \Delta r] d\omega }{ \hat u \cdot \int_{-\infty}^\infty \vec{S}(\vec{r}_f,\omega) d\omega },
\end{equation} 
The reshaping delay, as the very name suggests, arises from the deformation due to spectral modulation by the medium and can be calculated in terms of the spectral fields at the initial point as 
\begin{equation}
\Delta t_R = \wp \{ \exp[-\mathrm{Im}(\vec k) \cdot \Delta \vec{r}] \vec{E}(\vec{r}_i, \omega)\} \wp \{\vec{E}(\vec{r}_i, \omega)\},
\end{equation}
where the operator 
\[ \wp \{\vec{E}(\vec{r}, \omega)\} =\frac{\hat u \cdot \int_{-\infty}^\infty \mathrm{Re} [-i \frac{\partial\vec{E}(\vec{r}, \omega)}{\partial \omega}\; \times\; \vec{H}^*(\vec{r},\omega)] d\omega  }{ \hat u \cdot \int_{-\infty}^\infty \vec{S}(\vec{r},\omega) d\omega }, \]
This approach which is applicable to an arbitrary waveform or dispersive medium showed that both the components in general were always significant. The most important aspect of this proposal is that it does not involve any perturbative expansion of the wave number around the carrier frequency. A few salient points may be noted in this context: \begin{itemize}
\item For a narrowband pulse, the reshaping delay tends to zero and the total delay time is dominated by the group delay time. We note that even the Wigner delay time would describe the situation quite well in this case.
\item  For a broadband pulse with only propagating components, the net group delay can become negative, but the corresponding reshaping delay causes the overall delay time  to remain luminal as the pulse experiences a strong reshaping during propagation. This makes this proposal a very strong candidate to represent the traversal time that is causal and non-negative. 
\item It has been shown that the definition is equivalent to another definition based on the rate of energy absorbed by a detector 
\[ \langle t \rangle_r = \frac{ \int_{-\infty}^\infty t \frac{d A(\vec{r},t)}{dt}  dt}{  \int_{-\infty}^\infty \frac{d A(\vec{r},t)}{dt} dt}, \]
where$(dA/dt)$ is the rate of absorption of energy per unit volume inside the detector placed at $\vec r$, and is given by 
\[ \frac{dA}{dt} = \int \int \omega \left[ \varepsilon_0 \mathrm{Im}(\varepsilon) E^*(\omega') E(\omega) + \mu_0\mathrm{Im}(\mu)H^*(\omega) H(\omega) \right]
 d\omega'~d\omega. \]
This is integrated spatially over the detector volume to obtain the total rate of absorption within the detector. Obviously the spatial extent of the detector is assumed to be small compared to the length scales of propagation or spatial pulse widths.
\item In a medium such as a plasma, however, it has been shown~\cite{lipsa_pre} that a positivity of the traversal time in not obtained, particularly for broad-band pulses, which have large amounts of evanescent frequency components when the carrier frequency is smaller than the plasma frequency. This is related to the Hartmann effect~\cite{hartmann}, whereby the time for tunelling at far-sub-barrier energies becomes almost constant and negates the possibility of using this definition in principle for all situations.  Nevertheless, it should be noted that it is still operationally one of the most useful definitions of the traversal times and has been validated in experiments involving both temporally dispersive and angularly dispersive situations. 
\end{itemize}

\section{Quantum clocks {\label{sec:IV}}}
Due to issues with the various timescales which fail to conform to our intuitive understanding of the traversal timescales, attempts were made to develop``quantum clocks''  that measure the time the particle spends in a given volume. In analogy with a classical clock, the clock should tick only when the particle is within the region of interest. This is accomplished by coupling other degrees of freedom to other fields localized within the region of interest. Dynamical evolution of those degrees of freedom occur only when the particle is present in the region of interest. Thus, the expectation values of quantities associated with those degrees of freedom will translate to expectation values of the times spent in the region of space. While clocks have been proposed in principle for a long time~\cite{salecker}, three powerful methods that can have direct experimental implementation have been proposed and are popular. We will discuss the main ideas behind these proposals.

\subsection{B\"uttiker-Landauer oscillating barrier times}

B\"{u}ttiker and Landauer proposed~\cite{buttiker_landauer_prl} that the time of traverse of a charged particle through a potential barrier could be timed by super-imposing a time-harmonic electromagnetic field on top of the potential only within the region of interest. Suppose  is the region in which we seek the time of sojourn, the potential would be $V_0(\vec r) + V_1(\vec r) \cos (\omega t)$ where $V_0$ is the original static potential and the perturbing oscillating field has a magnitude ($V_1$)  that is constant within the region of interest and is zero outside (see Fig.~\ref{fig4} for a schematic depiction).  Interaction of the charged particle with the electromagnetic field would cause cause the absorption or emission of quanta of radiation. Thus, if the oscillation frequency was very small, the energy broadening of the transmitted or reflected spectrum would not be visible. The particle effectively sees a static potential and the transmission and reflected fluxes will adiabatically vary in time with the potential barrier height that changes harmonically. Thus, the times of interaction or traversal are very small compared to the time period of the oscillation. On the other hand, at high frequencies of oscillation, the particle would see the potential undergoing many oscillations during the time of its stay in the region and it would exchange quanta with the field. Then the energy spectrum of the reflected or transmitted particle would have energy sidebands separated from the incident energy by the energy of the quanta ($\pm \hbar \omega$). In this limit, the period of the electromagnetic field is clearly much smaller than the traversal time ($\tau_s$). There would be a crossover  between the two regimes of low or high frequencies and the period of the electromagnetic field during the crossover regime ($\omega \tau_s \simeq 1$) would a good measure of the time of traversal, regardless of whether the quantum wavefunction has a propagating nature or evanescent nature. 

\begin{figure}[t]
\includegraphics[width=0.9\linewidth]{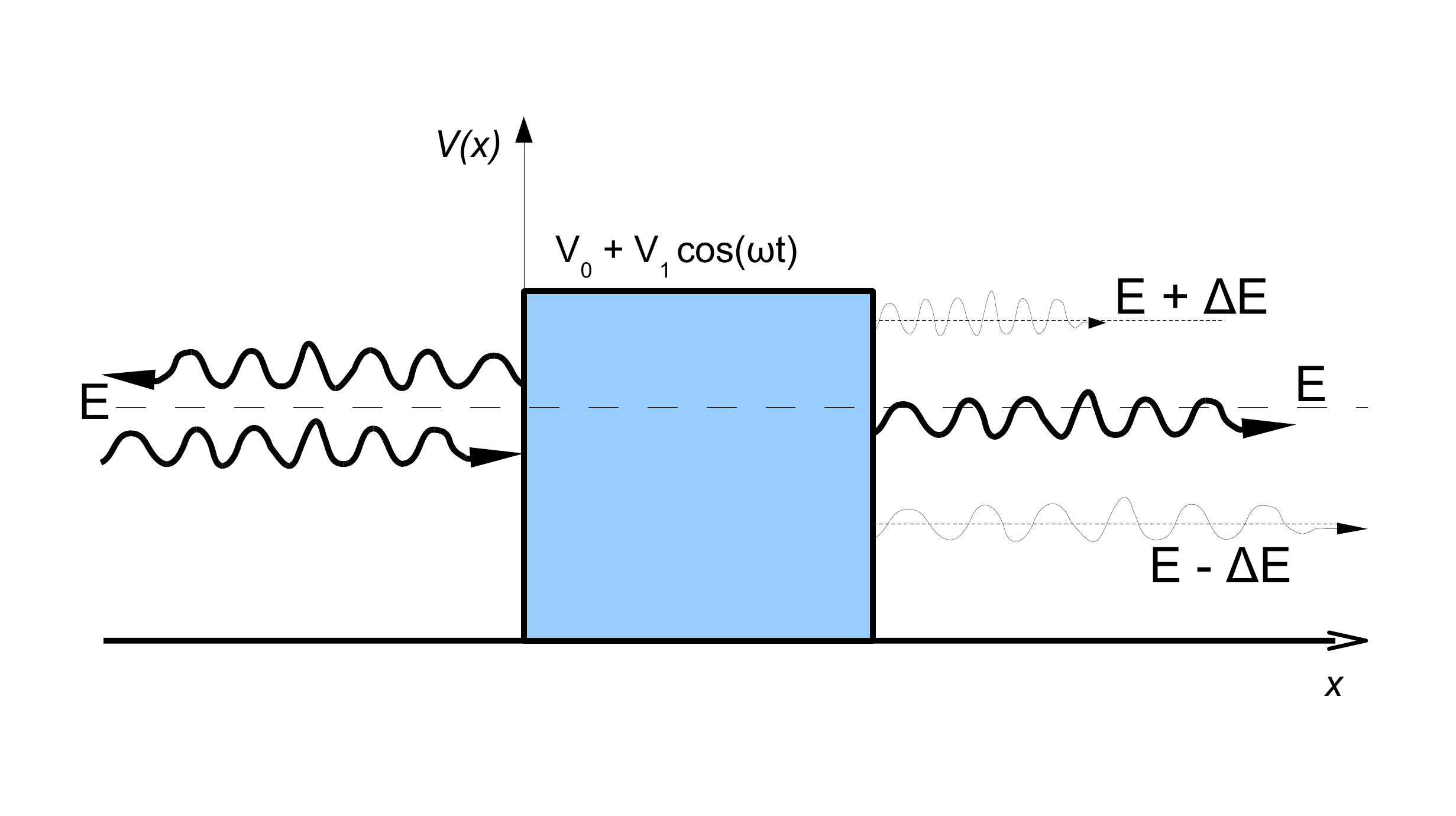} 
\caption{Schematic picture  depicted the tunnelling of a particle through a potential barrier with a time dependent strength. The particle exchanges quanta of energy (photons) with  the radiation field and the transmission develops energy sidebands. The particles with higher energy would tunnel through more efficiently than the particles with lower energy.  The period of the potential oscillations when the energy sidebands develop gives a timescale for the traversal of the particle through the barrier}
\label{fig3}
\end{figure}

In the one-dimensional case of a particle tunneling through a rectangular potential barrier, Buttiker and Landauer~\cite{buttiker_landauer_prl} obtain a tunneling time of
\begin{equation}
\tau_{BL} = \left[ \frac{m}{2(V_0-E)} \right]^{1/2} d, 
\end{equation}
where $E$ is the energy of the tunneling particle and $d$ is the width of the barrier. Using the WKB approximation at energies well below the barrier height, this was further generalized to  spatially varying potential barrier in one-dimension as 
\begin{equation}
\tau_{BL} = \int_{x_1}^{x_2}  \left[ \frac{m}{2(V_0-E)} \right]^{1/2} dx =   \int_{x_1}^{x_2} \frac{m}{\hbar \kappa(x)} dx, 
\end{equation}
where $\hbar \kappa(x) = \sqrt{2m(V_0(x) - E)}$   behaves as the instantaneous momentum of the tunneling particle. For more general potential shapes and energies, the calculations of the traversal times by this approach become very difficult. This approach, however, clearly sets out a timescale for the problem, particularly for the limiting case of low energy tunneling,  which any other valid  approach would need to reproduce.  An approach where the flow of the particle through the barrier region was constructed in terms of two counter-propagating  streams within the WKB approximation also obtained the traversal times consistent with the one obtained here~\cite{jayan_pramana}.

\subsection{B\"{u}ttiker's spin clock (Larmor precession and spin flip)}

\begin{figure}[t]
\includegraphics[width=0.8\linewidth]{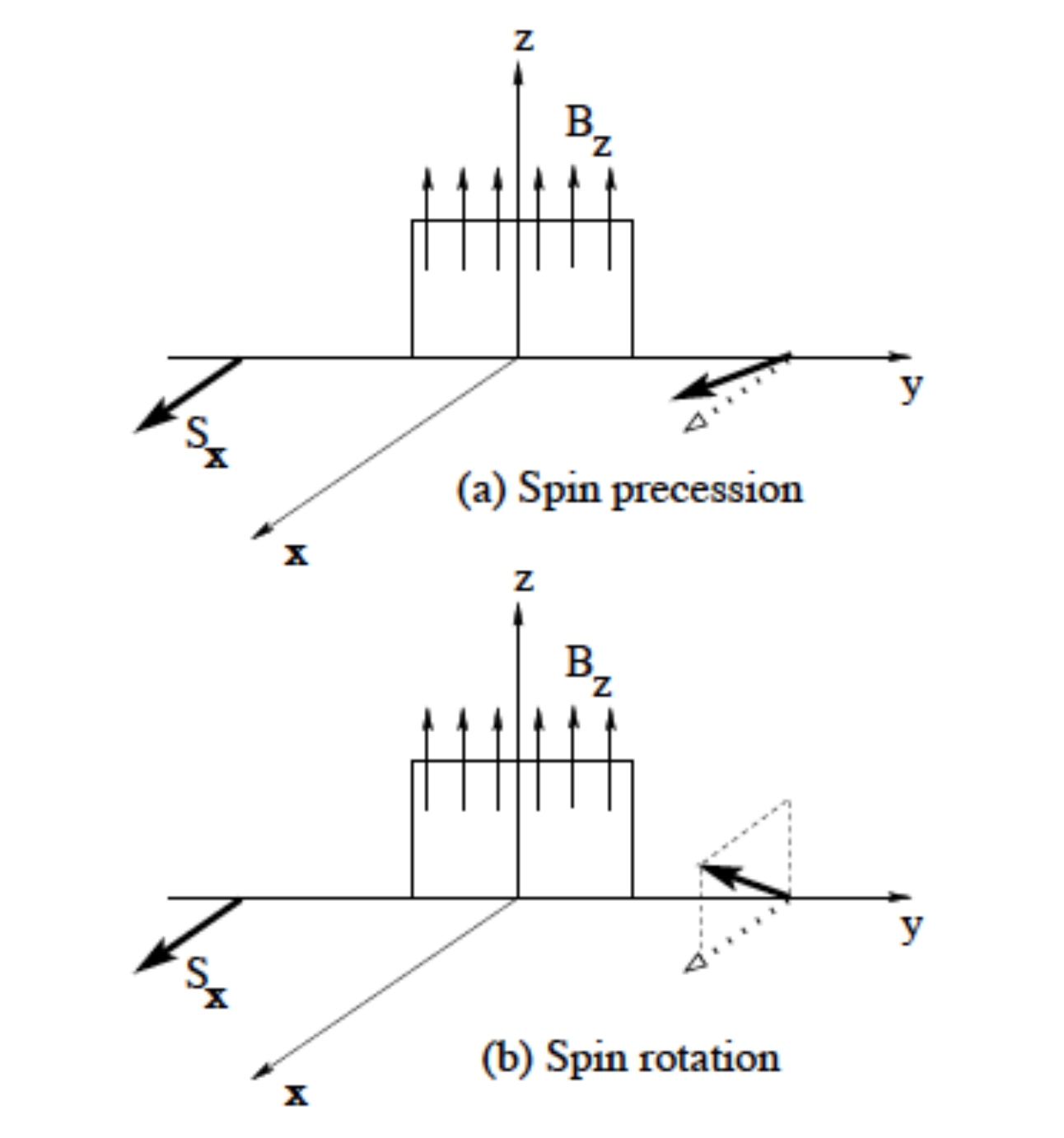} 
\caption{Schematic picture  depicting the  precession and spin rotation of the spin associated with a particle moving through a potential barrier where a  magnetic field is applied locally. The extent of the precession and rotation in the limit of a small magnetic field can yield the time spent by the particle within the region of the magnetic field. }
\label{fig4}
\end{figure}

A second clock related to using the Larmor precession of a quantum particle with an associated magnetic moment (or spin), such as an electron or a neutron, in an applied magnetic field (B). The rate of the precession of the spin $\omega_L = g\mu_B B/\hbar$,  ($\mu_B$ is the Bohr magneton) is constant in a spatially constant  magnetic field and could, thus, act as a possible clocking mechanism.  Consider a  magnetic spin that is initially polarized along the x-axis and moving along the y-axis through a region with potential, $V(y)$, wherein a uniform magnetic field along the z-axis is applied (see Fig.~\ref{fig4} for a schematic depiction).  If the particle takes a time ($\tau_y$) to traverse through the region of the magnetic field, the spin components of the transmitted flux would be $\langle S_y\rangle = -(\hbar/2) \omega_L\tau_y$ to the lowest order in the magnetic field.  Thus, measurement of the spin precession could yield the traversal time as
\begin{equation}
\tau_y = \frac{2}{g\mu_B} \lim_{B\rightarrow 0} \frac{\partial \langle S_y\rangle}{\partial B}
\end{equation}
B\"{u}ttiker~\cite{buttiker_larmor} recognized that the spin has a tendency to align along the magnetic field direction that he called {\it spin rotation} in addition to the Larmor precession in the place perpendicular to the magnetic field ({\it spin precession}).  Thus, there could be two time scales associated with the extents of the spin rotation ($\tau_z$)   and the spin precession ($\tau_y$).  The Hamiltonian for the case of a rectangular barrier is 
\begin{equation}
H = \left\{ \begin{array}{l} (p^2/2m + V(y)) I - (\hbar \omega_L/2) \sigma_z ~~~\forall~~~ 0 < y < L,  \\    p^2/2m I~~~~\mathrm{elsewhere} \end{array} \right.
\end{equation}
where $V(y)$ is the spatially constant potential, $I$ is the identity matrix and $\sigma_z$  is the z-component of the Pauli spin matrix and $H$ acts on the spinor $ \psi = \left( \begin{array}{c} \psi_+(y) \\ \psi_-(y) \end{array} \right)$, where $\psi_\pm$ are are the Zeeman components that represent the anti-parallel and parallel spin amplitudes.  The input spinor for a particle polarized along the x-axis can be written as equal superpositions of these components outside the region of the magnetic field due to which the expectation value of the Sz component would be zero. In the presence of the magnetic field , however, the kinetic energy for these components differ by the Zeeman energy of $\pm \hbar\omega_L/2$, and this gives a different value to the wave-vectors of the two components in the potential region. This is particularly severe in the case of tunneling at energies $(E)$ below the barrier when the exponential decay for the wavefunctions becomes very different:
\begin{equation}
\kappa_\pm = \left\{ \frac{2m}{\hbar^2} (V_0 -E) \mp \frac{m\omega_L}{\hbar} \right\}^{1/2} \simeq \kappa \mp \frac{m\omega_L}{2\hbar \kappa}, 
\end{equation} 
where $\kappa^2 = 2m(V_0-E)/\hbar^2$.  Thus, one spin component has a  greater probability to tunnel across than the other and the transmitted flux becomes spin polarized. 

The transmittance amplitudes for the two spin components can be approximately calculated in the case of energies far below the barrier height as $T_\pm \simeq T \exp(\pm \omega_L\tau_z)$, where $\tau_z = ml/\hbar \kappa$. The expectation values of the z-spin component of the transmitted flux is easily obtained as 
\[ \langle S_z \rangle = \frac{\hbar}{2} \frac{T_+ - T_-}{T_++T_-} = \frac{\hbar}{2} \tanh (\omega_L\tau_z) \simeq \frac{\hbar}{2} \omega_L\tau_z \]
where the last approximation is made in the limit of small magnetic fields. Thus, the z-component of the spin scales linearly with the magnetic field and presents yet another clocking mechanism and the spin rotation time can be defined as
\begin{equation}
\tau_z = \frac{2}{g\mu_B} \lim_{B \rightarrow 0} \frac{\partial \langle S_z\rangle}{\partial B}
\end{equation}
It turns out that the spin precession ($S_y$)  dominates for energies far above the barrier and the spin rotation ($S_z$) dominates for energies far below the barrier.For a rectangular barrier, the spin precession time for energies far above the barrier tend to the Wigner delay times and far below the barrier, the spin rotation times tend to the B\"uttiker-Landauer times for the oscillating barrier.  B\"uttiker proposed that a net traversal time could be defined as the Pythogorean sum of the two spin times as $\tau^2 = \tau_y^2  + \tau_z^2$, which reflects the vectorial nature of the change of the spin. This prescription does not, however, have any fundamental basis. 

\subsection{Absorption and amplification as a clock}
A third interesting manner to clock the time of sojourn in a given region of space would be absorption or amplification, which would be at a rate proportional to the amplitude of the wave. Hence, a measurement of the wave amplitude as it enters and exits the region of interest can yield the time that it spends inside. Since the growth is exponential, a logarithmic derivative of the transmittance / reflectance with respect to the imaginary potential is needed.  For light, both absorption and stimulated emission are coherent processes that leave the phase of a coherent mode unchanged, due to the Bosonic nature and this definition in terms of absorption or amplification is a natural definition. Such a process would not be strictly applicable to Fermionic particles like electrons or neutrons. However, the effects of absorption or amplification would eventually need to be discussed in the limit of infinitesimally small levels of absorption/amplification that tends to zero, so that the original problem remains unchanged. Hence, we may assume that the formal procedure works for Fermionic particles as well. For a scalar wave, coherent absorption / amplification may be implemented by adding an imaginary potential ($iV_I$) only in the region of interest (instead of the magnetic field as in Fig.~\ref{fig4}) , which will eventually be made zero ($V_I \rightarrow 0$). The Schrodinger wave equation for the wave function of a particle in the presence of the imaginary potential becomes
\begin{equation}
i\hbar \frac{\partial \psi}{\partial t} = -\frac{\hbar^2}{2m} \nabla^2 \psi + [ V_0(\vec r) + i V_I] \psi.
\end{equation}
Then the traversal times for the transmission and reflection can be defined as
\begin{equation}
\tau^{(T)}  = \frac{\hbar}{2} \lim_{V_I \rightarrow 0} \frac{\partial \ln \vert T \vert^2 }{\partial V_I} , ~~\mathrm{and}~~ 
\tau^{(R)}  = \frac{\hbar}{2} \lim_{V_I \rightarrow 0} \frac{\partial \ln \vert  R\vert^2 }{\partial V_I}
\end{equation}
respectively, where T and R are the complex transmission and reflection coefficients. 

\begin{figure}[t]
\includegraphics[width=0.8\linewidth]{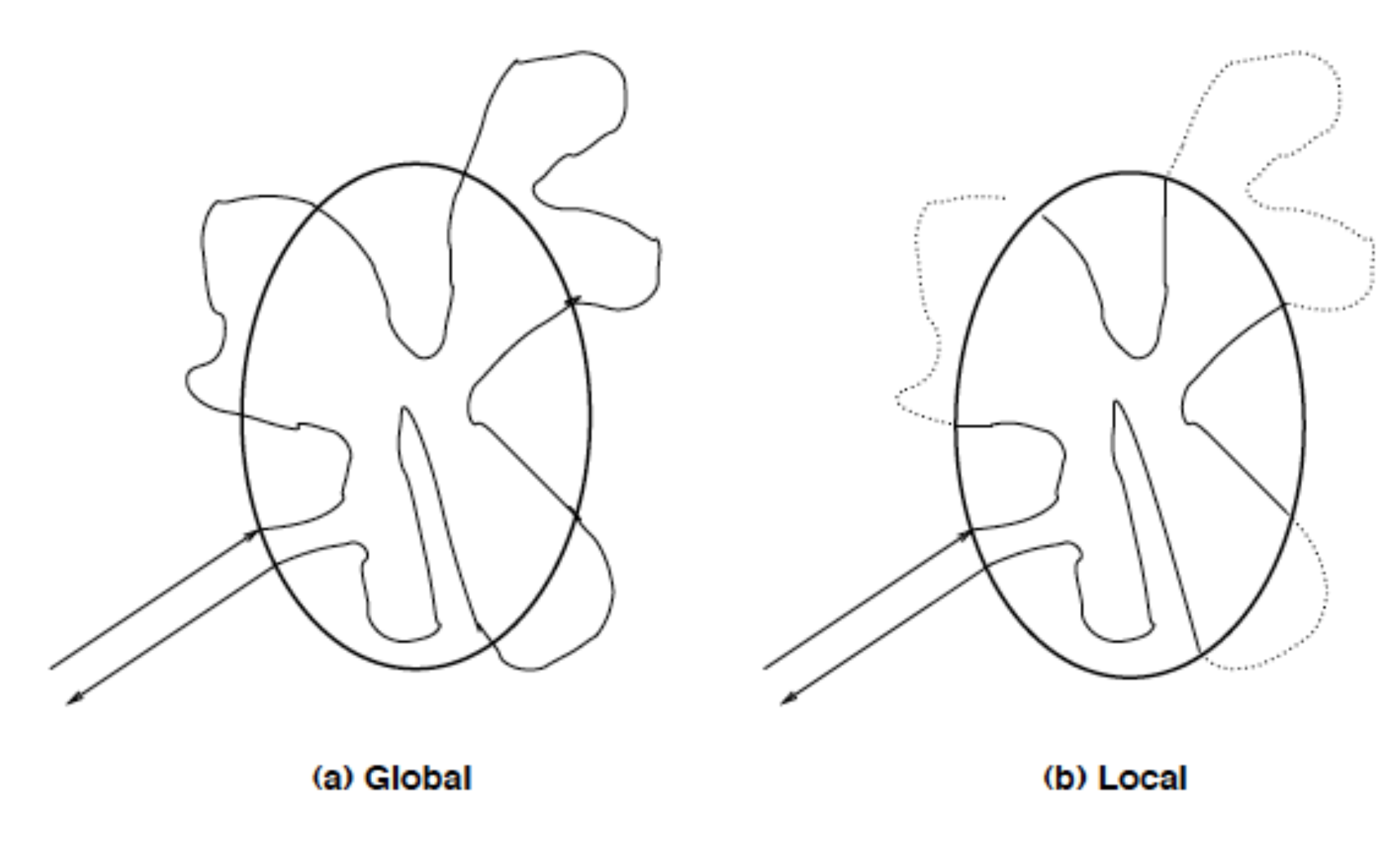} 
\caption{Schematic diagram  depicting a possible pathway in the Feynman path integral sense. (a) shows the portion affected by a global variation of the potential, and (b) shows the portion (solid line) affected by a local variation of the potential. The dotted portion of the path is not affected and should not be counted towards the traversal time.}
\label{fig5}
\end{figure}

The idea was originally suggested by by Pippard to Buttiker~\cite{buttiker_pvt}, who found that while it reproduced the spin precession times, it did not recover the spin rotation time for waves that are evanescent in the region of interest (sub-barrier tunneling). It reproduces the Wigner delay times in the limit of large energy above the barrier height.  This identity can be understood in terms of the analytic properties of the complex reflection and transmission coefficients. Let us consider the transmission coefficient in the complex energy plane $(E_r, E_i)$. Its logarithm
$\ln (T) - \ln \vert T \vert + i \mathrm{Arg}(T)$ would be an analytic function within the Riemann sheet and would satisfy the Cauchy Riemann conditions:
\[ \frac{\partial \ln \vert T \vert}{\partial E_r} =  -  \frac{\partial \mathrm{Arg}(T) }{\partial E_i }, ~~\mathrm{and} ~ ~
\frac{\partial \ln \vert T \vert}{\partial E_i} =    \frac{\partial \mathrm{Arg}(T) }{\partial E_r }. \]
Thus, for energies far above the barrier, the left hand side of the second relation is directly proportional to the traversal times obtained by the imaginary potential ($E_i = V_I$), while the right hand side relates directly to the kinetic energy of the particle as potential energy is comparatively small ($E_r \gg V_0$). This is why the traversal times tend to the Wigner delay time for large energies. A similar argument will hold for the reflection delay times as well.  Note, however, that it is assumed that the potentials are varied everywhere in space in these arguments.  Usually for the a quantum clock, the potential should only be varied locally within the region of interest (see Fig.~\ref{fig5}). 
	The issues with the traversal times obtained by the imaginary potential clocks refuse to remain positive for all potentials and further do not tend to the  Büttiker-Landauer times for energies far below the barrier. For the case of tunneling across a rectangular barrier, the ratio $\tau^{(T,R)} / \tau_{BL} \rightarrow 0$ in the low energy limit. The  Büttiker-Landauer times are very compelling in that limit and we  would really want the traversal to remain positive definite. 

\section{ Sojourn times: correcting the clocks}
It turns out that most of the problems of defining a positive definite traversal time for the transmission arise from spurious scattering concomitant with the very clocking mechanism / potential  and a procedure was duly outlined  to separate out this extra scattering and correct the traversal times~\cite{sar_epl}. We will call these corrected traversal times as sojourn times  as they literally relate to the times of journey through the region of interest.  We will primarily deal with the imaginary potential clock here, but will point out the exact analogies with the Larmor clock.
\begin{figure}[t]
\includegraphics[width=0.8\linewidth]{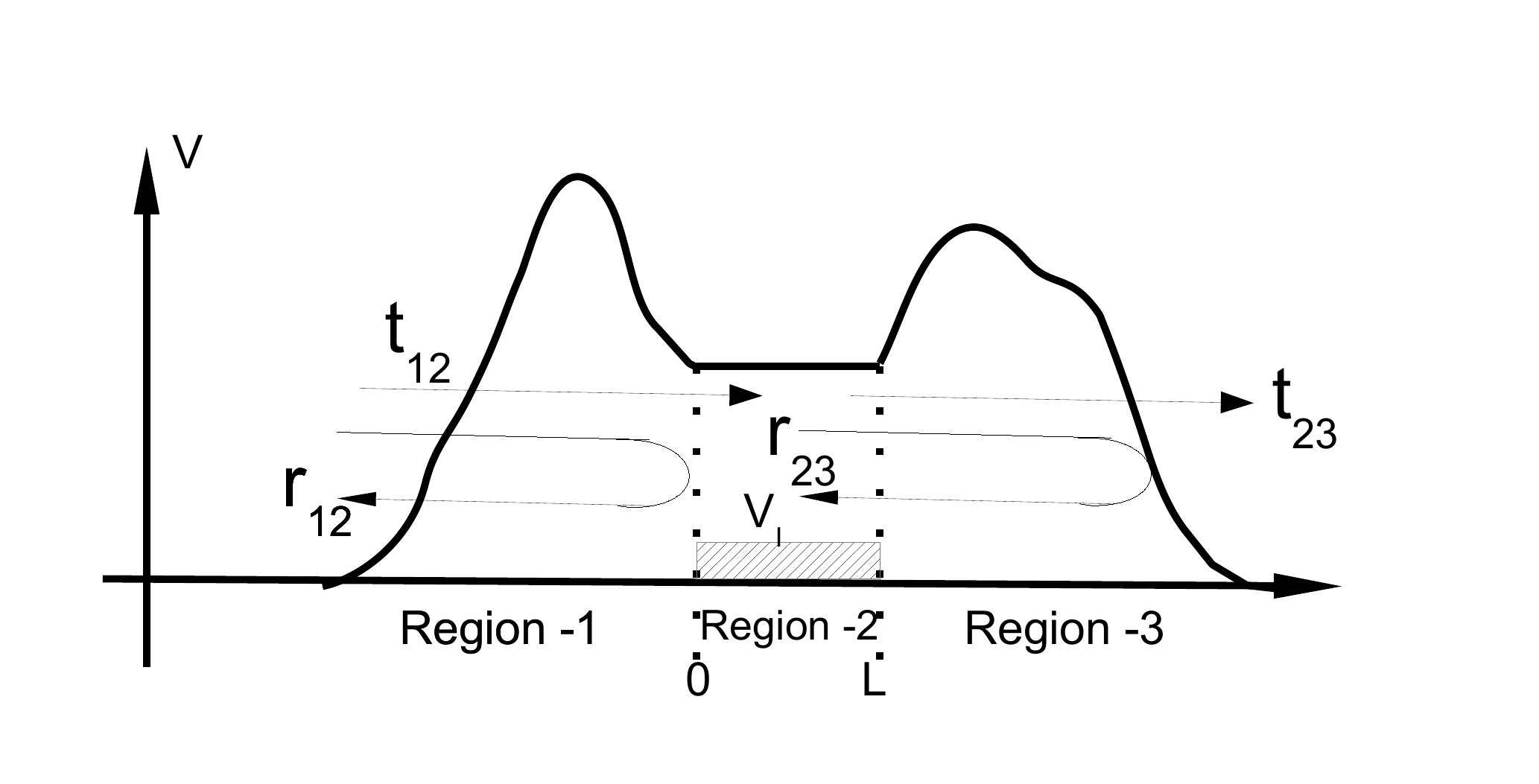} 
\caption{Schematic diagram  partial transmission and reflection coefficients associated with the potentials (regions 1 and 3) on either side of a constant potential region (2). These partial transmission($t_{jk}$ and reflection ($r_{jk}$) coefficients may be used to construct the net transmission and reflection coefficients using equations (~\ref{eqn:partial_waves}). }
\label{fig6}
\end{figure}
	First of all, it is important to appreciate that the presence of the clock potential not only invokes a response in the relevant extra degree of freedom, but also modifies the scattering due to a change of potential. One would expect this extra scattering to go to zero in the limit of zero clock potential. Our procedure, however, involves taking a derivative with respect to the clock potential, which can have contributions that would not vanish as the clock potential is made zero.  Let us consider the transmission of a wave with energy above the barrier (case of propagating waves) through the constant potential region encased between two arbitrary potentials on either side as shown in Fig.~\ref{fig6}. The space may be divided into three regions and the partial coefficients of transmission and reflection for a wave incident from region (j) unto region (k) are $t_{jk}$ and $r_{jk}$ as shown schematically in Fig.~\ref{fig6}. The transmission and reflection coefficients can be written as a sum of the partial waves through the region~\cite{born_wolf}
\begin{eqnarray}
T&=& t_{12} t_{23} e^{ik'L} + t_{12} r_{23} r_{21} t_{23} e^{3ik'L} + t_{12} r_{23} r_{21} r_{23} r_{21}t_{23} e^{5ik'L} + \cdots  \nonumber \\
R&=& r_{12} +  t_{12} r_{23}  t_{21} e^{2ik'L} + t_{12} r_{23} r_{21} r_{23} t_{21} e^{4ik'L} +\cdots
\label{eqn:partial_waves}
\end{eqnarray}
the coefficients $t_{jk}$ and $r_{jk}$  also depend on the clock potential and a derivative with respect to the total transmission or reflection coefficient would leave behind some terms arising from this dependence, which do not vanish as the clock potential is made to go to zero. This is the origin of the spurious scattering.  An analysis of the structure of the partial wave superposition quickly reveals that the growth or attenuation related to the imaginary potential would only involve the paired combination of ($V_I L$) where $L$ is the lenght of the spatial region of interest.  Afterall the traversal time should directly relate to spatial region of interest. The spurious scattering on the other hand would only involve unpaired $V_I$.  A formal procedure to isolate the effects of this spurious scattering can now be given. We will treat$\xi = V_I L $ and $V_I$ as independent variables, keep $\xi$ formally constant and let $V_IL \rightarrow 0$ in the expression for the transmission coefficient.  The sojourn time for transmission for propagating waves is now obtained as 
\begin{equation}
\tau_s^{(T)} = \frac{\hbar L}{2} \lim_{\xi \rightarrow 0} \frac{\partial \ln \vert T(\xi, V_I=0) \vert^2}{\partial \xi} .
\end{equation} 
For the case of wave tunneling (energy lesser than the barrier height $E < V_0$), the wave vector is principally imaginary in the barrier region. The real part of the potential or refractive index essentially affects the rate of exponential decay / growth of the evanescent wave. On the other hand, the imaginary potential or imaginary part of the refractive index causes a phase shift with distance of this evanescent wave~\cite{lipsa_pre}.  Consider the complex wave-vector for propagating waves ($E > V_0$),
\[ k = \sqrt{ \frac{2m}{\hbar^2} [E - (V_0+iV_I)]} ~~\simeq k_r - i \frac{mV_I}{k_r \hbar^2} ~~ \forall ~~ V_I \ll E,   \]
where $\hbar k_r = \sqrt{2m(E-V_0)}$. When we have evanescent waves ($E < V_0$) on the other hand,  we can write 
\[ k = \sqrt{ \frac{2m}{\hbar^2} [E - (V_0+iV_I)]} ~~\simeq i \kappa_r -  \frac{mV_I}{\kappa_r \hbar^2} ~~ \forall ~~ V_I \ll  V_0, \]
where $\hbar \kappa = \sqrt{2m(V_0-E)}$.We hence realise the principle effect of the clock with respect to the spatial region is in the phase of the wave and not the amplitude for the evanescent wave. This is completely analogous to the spin rotation being predominant over the spin precession in the case of the Larmor clock. Mathematically, we have a square root singularity for the wave-vector in the complex energy plane and we are unable to analytically continue the behaviour for propagating waves to the case of evanescent waves across the branch-cut in either case of the imaginary potential clock or the Larmor clock. Hence, we define for the case of sub-barrier tunneling or evanescent wave, the sojourn time as
\begin{equation}
\tau^{(T)}_s  = \frac{\hbar L}{2} ~ \lim_{\xi \rightarrow 0}~\frac{\partial}{\partial \xi} \left[ \ln  \left(\frac{T(\xi, V_I=0)}{T^*(\xi, V_I=0)} \right)  \right]  ,
\end{equation}
where $T*$ is the complex conjugate of $T$ and the ratio $T/T*$ essentially yields twice the phase of the transmission coefficient. 
	For the case of the region with constant potential of height $V_0$ enclosed between the two arbitrary potentials and energy above the barrier, the sojourn time  reduces to a positive definite quantity:
\[ \tau^{(T)}_s  =  \frac{1 - \lvert r_{21} r_23 \rvert^2 }{1 + \lvert r_{21} r_{23} \rvert^2 - 2\mathrm{Re}(r_{21} r_{23} e^{2i k_r L})} \tau_{BL}.  \]
A similar positive definite expression is obtained in the case of evanesent waves (tunneling below the barrier) as
\[ \tau^{(T)}_s  =  \frac{1 - \lvert r_{21} r_23 \rvert^2 e^{-4\kappa L} }{1 + \lvert r_{21} r_{23} \rvert^2e^{-4\kappa L} - 2\mathrm{Re}(r_{21} r_{23} e^{-2\kappa L})} \tau_{BL}.  \]
The sojourn time tends directly to the B\"uttiker-Landauer times for an opaque barrier (energies far below the barrier  or long barrier lengths), which is very important. In the high energy limit, it tends to the classical Wigner delay times. Thus it has the correct limits. Further, it was shown in Ref.~\cite{sar_epl} that the sojourn times defined in this manner  for two non-overlapping regions is additive and hence, the times for any region of space with any arbitrary applied potential can be concluded to be positive definite. 

	The case of reflection is a little bit more complex.  Applying the above definitions for the reflection coefficient in place of the transmission coefficient does not yield a positive definite sojourn time.  From the expansion in terms of the partial waves, one notes an essential difference between the reflection and the transmission. All the partial waves for the transmitted wave sample the region of interest and pick the paired combination $\xi = V_IL$ in the amplitude or the phase. In the case of reflection, there is one partial wave $r_{12}$ corresponding to the prompt reflection from the edge of the region of interest that never enters the region of interest. Yet, this partial wave interferes with the others to produce the net reflection amplitude.  In the  spirit of our earlier arguments that the sojourn time should be causally related to the region of interest, it would be necessary to eliminate the weightage of this partial wave that should not be affected by the clocking potential.  Hence, we explicitly subtract this prompt reflection amplitude out of the total reflection amplitude $R' = R -  r_{12}$  and define the sojourn times for reflection using $R'$ in place of $T$ as before for both cases  of the propagating waves as well as the evanescent waves. A simple and general result is obtained as
\begin{equation}
\tau^{(R)}_s = \tau^{(T)}_s + \tau_{BL}
\end{equation}
which is consequently always greater than the sojourn time for transmission and  positive definite. An experimental measurement of this procedure is possible by interfering destructively the reflection from a modified potential whose reflection is  $r_{12}$ with the reflectance from the  given potential. For example, one can use the same optical system that is index matched to the continuum form beyond the point where the absorption or gain is applied. Alternatively, one may use the recently developed metamaterial perfect absorbers~\cite{govind_sar} for light, where there is the possibility of matching the impedance and preventing any reflection. 

It should be noted that such a corection procedure can be analogously applied to the spin precession and spin rotation times for the Larmor clock as well, where the paired variable $\xi = BL$ would be taken to zero after explicitly putting the magnetic field $B=0$ while keeping $\xi$ formally constant. We then  obtain identical results to the imaginary potential clock. The procedure outlined above was shown equivalent to stochastic absorption~\cite{jayan_stochastic},  whereby the absorption does not cause any scattering but contributes only to loss of the wave flux. This can be viewed as the case when there is inelastic scattering out of the given mode of the mesoscopic and the coherence of the mode is not affected.  Only the coherent part of the wave is measured and the scattering into other modes manifests as a loss for this mode.  Another example could be that the scattering leads to decoherence of the wave. In any interferometric measurement only the coherent part of the wave is measured and the decohered part would appear as a loss. Thus, a finite rate of decoherence itself could be utilized as a clocking mechanism . 

\subsection{The problem of the quantum first passage time}
Consider a free quantum particle that is released from a localized region at some instant of time and thereafter subjected to instantaneous projective measurements to detect its arrival at a particular region of space. The  measurements are made at regular time intervals  , and the system is allowed to evolve until the time a detection occurs. The main question addressed is: what is the probability that the particle is detected for the first time after time $t$ i.e. at the $n=(t/\tau)^{th}$ measurement~\cite{nkumar_pramana}. Conversely, one can ask for the probability of particle not being detected (i.e., surviving) upto a given time.

It is proposed by a general perturbative approach for understanding the dynamics which maps the evolution operator, which consists of successive unitary transformations followed by projections, to one described by a non-Hermitian Hamiltonian. For some examples of a particle moving on one- and two-dimensional lattices with one or more detection sites, use this approach to find exact expressions for the survival probability and find excellent agreement with direct numerical results. For the one- and two-dimensional systems, the survival probability is shown to have a power-law decay with time, where the power depends on the initial position of the particle. It is shown that an interesting and nontrivial connection between the dynamics of the particle in their model and the evolution of a particle under a non-Hermitian Hamiltonian with a large absorbing potential at some sites~\cite{adhar15, dhar_pra}. If continuous projective measurements are done then famous Zeno comes in to effect and particle does not evolve. Thus quantum first passage time indeed a nontrivial model. It is very subtle and as mentioned before in quantum mechanics there is no dynamical operator for time travel between two points.
We may conclude this brief disussion on this topic by saying  that the quantum first passage time is still remains a mystery

\section{Conclusions and Outlook {\label{sec:VIII}}}
We have outlined here various attempts to answer a fundamental question, namely, “what is the time that a quantum particle or wave spends in a specified region of space ?”  We do not have a self-adjoint operator for the arrival time in quantum mechanics, and the arrival time is not an observable.  Yet it is intuitive and important to ask about timescales of any physical problem and the time of stay or sojourn appears to be a calculable quantity and a useful one to compare timescales. The sojourn time  can provide for a meaningful alternative view-point within quantum mechanics. Knowledge of the sojourn times  obviously cannot provide answers to questions that cannot be answered within the paradigm of quantum mechanics. Here we have exploited the fact that light as well as quantum particles are mathematically described by the same Helmholtz equation and talk about both systems in the same breath.   
	Starting with the description of wavepackets by the group velocity and its extension by Wigner, we have indicated a variety of manners in which this question has been approached. The Smith dwell time is a positive definite time, but it is an unconditional time that does not depend on the output scattering channel. The path integral approach~\cite{nkumar_pramana} or an approach based on the WKB method~\cite{jayan_pramana} are  closely related to this, but can define conditional traversal times. Beyond these, we discuss the proposals to consider the dynamical evolution of an extra degree of freedom attached to the traversing particle due to the local interaction with a potential applied only in the region of interest. One has to meaningfully identify the extent of evolution of the clocking mechanism, which is an observable, with the time spent in that region. We have discussed three examples:
(i) exchange of  quanta via interaction with a radiation field;
(ii) the spin precession; 
(iii) rotation in a magnetic field or the growth / attenuation due to an imaginary potential (amplifying or absorbing medium). 
A rather subtle problem that arises with these clocks is that they give rise to an extra spurious scattering that interferes with the very process and effectively changes the potential in the region of interest. This extra scatterings is carefully identified and is explicitly eliminated by a formal mathematical procedure. Further, it is shown that the effect of the clocking potential manifests in different quantities for propagating and evanescent waves: In the case of the imaginary potentials, it manifests in the amplitude for propagating waves and in the phase for evanescent waves (tunneling below the barrier); and in the case of the Larmor spin clock, it mainfests in the spin precession for propagating waves and in the spin rotation for evanescent waves. As a final caveat, it is shown that partial waves that reflect from the surface of the region should also be eliminated from reckoning as they do not spend anytime within the region of interest and this can, indeed, be done by interferometric measurements. Including these considerations yields a sojourn time that is
 \begin{enumerate}
\item Real and positive definite; 
\item Additive for non-overlapping regions of space;
\item Related causally to the region of interest; 
\item Calculable and directly related  to a measurable quantity. \end{enumerate}

        The main issue  in defining traversal or sojourn times  for a quantum system has been due to interference between partial waves (the alternative paths in quantum mechanics), which defies naive realism. With these advances in understanding, however,  the sojourn time becomes a calculable quantity that is practically useful for estimating other quantities and understanding physical phenomena, for example, the dephasing rates in a quantum system.
\section{Acknowledgement}
Both the authors acknowledge illuminating and educative discussions with Prof. N. Kumar who introduced them to these topics as well as the constant  encouragement they have received from him. They would like to dedicate this tutorial to him. SAR acknowledges funding from the DST India through a Swarna Jayanti Fellowship. AMJ thanks DST, India for financial support (through J. C. Bose National Fellowship)

\end{document}